\documentclass[a4paper,11pt]{article}
\usepackage{pos}
\usepackage{tikz}
\usepackage{mathtools}
\usepackage{subcaption}
\usepackage{float}
\floatstyle{plaintop}
\restylefloat{table}
\usepackage{caption}
\usepackage{tikz}
\usepackage{pgfplots}
\usepackage{subcaption}

\renewcommand{\>}{\rangle}
\newcommand{\<}{\langle}
\newcommand{\cev}[1]{\reflectbox{\ensuremath{\vec{\reflectbox{\ensuremath{#1}}}}}}
\newcommand{\tmin}{\tau_{\text{min}}}


\title{Investigating the Compton amplitude subtraction function in lattice QCD}
\ShortTitle{Investigating the subtraction function in lattice QCD}

\manuallySeparateAuthors
\author*[a]{A.~Hannaford-Gunn,}
\author*[a]{ E.~Sankey,}
\author[a]{ K.~U.~Can,}
    \author[b]{ R.~Horsley,}
    \author[c]{ H.~Perlt,}
\author[d]{ P.~E.~L.~Rakow,}
    \author[e]{ G.~Schierholz,}
    \author[a]{ K.~Somfleth,}
      \author[f]{ H.~St\"uben,}
    \author[a]{ R.~D.~Young}
    \author[a]{ and J.~M.~Zanotti}
    \author{ for the {CSSM-QCDSF-UKQCD Collaboration}}

\affiliation[a]{CSSM, Department of Physics, University of Adelaide,
  Adelaide SA 5005, Australia}
  
\affiliation[b]{School of Physics, University of Edinburgh, Edinburgh EH9 3JZ, UK}
\affiliation[c]{Insitut f\"ur Theoretische Physik, Universist\"at Leipzig, 04103 Leipzig, Germany}

\affiliation[d]{Theoretical Physics Division, Department of Mathematical Sciences, University of Liverpool, Liverpool L69 3BX, UK}

    \affiliation[e]{Deutsches Elektronen-Synchrotron DESY, Notkestr.~85, 22607 Hamburg, Germany}
    
        \affiliation[f]{Regionales Rechenzentrum, Universit\"at Hamburg, 20146 Hamburg, Germany}

\emailAdd{alec.hannafordgunn@adelaide.edu.au}
\emailAdd{edward.sankey@adelaide.edu.au}

\abstract{Theoretical predictions of the proton--neutron mass
  difference and measurements of the proton's charge radius require
  inputs from the Compton amplitude subtraction function.
  Model-dependent and non-relativistic calculations of this
  subtraction function vary significantly, and hence it contributes
  sizeable uncertainties to the aforementioned physical quantities.
  We report on the use of Feynman-Hellmann methods in lattice QCD to
  calculate the subtraction function from first principles.
  In particular, our initial results show anomalous high-energy
  behaviour that is at odds with the prediction from the operator
  product expansion (OPE).
  Therefore, we investigate the possibility that this unexpected
  behaviour is due to lattice artifacts, by varying the lattice
  spacing and volume, and comparing different discretisations of the
  vector current.
  Finally, we explore a Feynman-Hellmann implementation
  that is less sensitive to short-distance contributions and show that
  the subtraction function's anomalous behaviour can be attributed to
  these short-distance contributions.
  As such, this work represents the first steps in achieving a
  complete understanding of the Compton amplitude subtraction
  function.  }

\FullConference{%
 The 38th International Symposium on Lattice Field Theory, LATTICE2021
  26th-30th July, 2021
  Zoom/Gather@Massachusetts Institute of Technology
}
\numberwithin{equation}{section}

\begin{document}
\maketitle

\section{Introduction}
\begin{figure}
    \centering
    \scalebox{0.5}{\tikzset{every picture/.style={line width=0.75pt}} 

\begin{tikzpicture}[x=0.75pt,y=0.75pt,yscale=-1,xscale=1]

\draw [line width=1.5]    (218,140) -- (103,142) ;
\draw [shift={(160.5,141)}, rotate = 179] [fill={rgb, 255:red, 0; green, 0; blue, 0 }  ][line width=0.08]  [draw opacity=0] (11.61,-5.58) -- (0,0) -- (11.61,5.58) -- cycle    ;
\draw [shift={(218,140)}, rotate = 179] [color={rgb, 255:red, 0; green, 0; blue, 0 }  ][fill={rgb, 255:red, 0; green, 0; blue, 0 }  ][line width=1.5]      (0, 0) circle [x radius= 4.36, y radius= 4.36]   ;
\draw [line width=1.5]    (452,142) -- (558,142) ;
\draw [shift={(505,142)}, rotate = 180] [fill={rgb, 255:red, 0; green, 0; blue, 0 }  ][line width=0.08]  [draw opacity=0] (11.61,-5.58) -- (0,0) -- (11.61,5.58) -- cycle    ;
\draw [shift={(452,142)}, rotate = 0] [color={rgb, 255:red, 0; green, 0; blue, 0 }  ][fill={rgb, 255:red, 0; green, 0; blue, 0 }  ][line width=1.5]      (0, 0) circle [x radius= 4.36, y radius= 4.36]   ;
\draw  [fill={rgb, 255:red, 155; green, 155; blue, 155 }  ,fill opacity=0.45 ][line width=1.5]  (218,142) .. controls (218,130.95) and (270.38,122) .. (335,122) .. controls (399.62,122) and (452,130.95) .. (452,142) .. controls (452,153.05) and (399.62,162) .. (335,162) .. controls (270.38,162) and (218,153.05) .. (218,142) -- cycle ;
\draw  [line width=1.5]  (140.63,42.19) .. controls (144.73,40.6) and (148.02,39.75) .. (149.14,41.02) .. controls (150.82,42.92) and (147.21,49.02) .. (143.42,55.41) .. controls (139.63,61.8) and (136.02,67.89) .. (137.7,69.79) .. controls (139.37,71.69) and (145.87,68.87) .. (152.68,65.9) .. controls (159.5,62.94) and (165.99,60.12) .. (167.67,62.02) .. controls (169.34,63.92) and (165.74,70.01) .. (161.94,76.4) .. controls (158.15,82.79) and (154.55,88.89) .. (156.22,90.79) .. controls (157.9,92.69) and (164.39,89.87) .. (171.21,86.9) .. controls (178.02,83.93) and (184.51,81.12) .. (186.19,83.02) .. controls (187.87,84.92) and (184.26,91.01) .. (180.47,97.4) .. controls (176.67,103.79) and (173.07,109.88) .. (174.75,111.78) .. controls (176.42,113.68) and (182.92,110.87) .. (189.73,107.9) .. controls (196.54,104.93) and (203.04,102.11) .. (204.71,104.01) .. controls (206.39,105.91) and (202.79,112.01) .. (198.99,118.4) .. controls (195.2,124.79) and (191.59,130.88) .. (193.27,132.78) .. controls (194.95,134.68) and (201.44,131.86) .. (208.25,128.89) .. controls (215.07,125.93) and (221.56,123.11) .. (223.24,125.01) .. controls (224.91,126.91) and (221.31,133) .. (217.52,139.39) .. controls (216.95,140.35) and (216.38,141.3) .. (215.84,142.23) ;
\draw  [line width=1.5]  (530.21,44.47) .. controls (527.52,43.44) and (525.46,43.07) .. (524.55,43.98) .. controls (522.77,45.78) and (526,52.07) .. (529.41,58.68) .. controls (532.81,65.28) and (536.05,71.58) .. (534.26,73.38) .. controls (532.47,75.17) and (526.16,71.97) .. (519.53,68.61) .. controls (512.91,65.24) and (506.59,62.04) .. (504.81,63.83) .. controls (503.02,65.63) and (506.26,71.93) .. (509.66,78.53) .. controls (513.07,85.14) and (516.3,91.44) .. (514.52,93.23) .. controls (512.73,95.03) and (506.42,91.83) .. (499.79,88.46) .. controls (493.17,85.09) and (486.85,81.89) .. (485.06,83.69) .. controls (483.28,85.48) and (486.51,91.78) .. (489.92,98.39) .. controls (493.32,104.99) and (496.56,111.29) .. (494.77,113.09) .. controls (492.99,114.88) and (486.67,111.68) .. (480.05,108.31) .. controls (473.42,104.95) and (467.11,101.75) .. (465.32,103.54) .. controls (463.53,105.34) and (466.77,111.63) .. (470.18,118.24) .. controls (473.58,124.85) and (476.82,131.14) .. (475.03,132.94) .. controls (473.24,134.74) and (466.93,131.54) .. (460.3,128.17) .. controls (453.68,124.8) and (447.36,121.6) .. (445.58,123.4) .. controls (443.79,125.19) and (447.03,131.49) .. (450.43,138.09) .. controls (451.5,140.18) and (452.56,142.22) .. (453.45,144.11) ;
\draw    (182,50) -- (224.06,99.71) ;
\draw [shift={(226,102)}, rotate = 229.76] [fill={rgb, 255:red, 0; green, 0; blue, 0 }  ][line width=0.08]  [draw opacity=0] (8.93,-4.29) -- (0,0) -- (8.93,4.29) -- cycle    ;
\draw    (440,102) -- (481.88,60.12) ;
\draw [shift={(484,58)}, rotate = 495] [fill={rgb, 255:red, 0; green, 0; blue, 0 }  ][line width=0.08]  [draw opacity=0] (8.93,-4.29) -- (0,0) -- (8.93,4.29) -- cycle    ;

\draw (150,149.4) node [anchor=north west][inner sep=0.75pt]  [font=\Large]  {$p$};
\draw (210,49.4) node [anchor=north west][inner sep=0.75pt]  [font=\Large]  {$q$};
\draw (441,52.4) node [anchor=north west][inner sep=0.75pt]  [font=\Large]  {$q$};
\draw (507,145.4) node [anchor=north west][inner sep=0.75pt]  [font=\Large]  {$p$};

\end{tikzpicture}}
    \caption{Forward Compton scattering for a proton with momentum $p$ and virtual photon with momentum $q$.}
    \label{fig:1}
\end{figure}
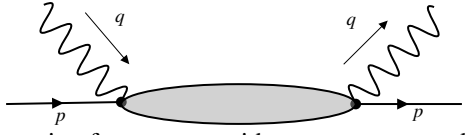



The forward Compton amplitude is a necessary input into two important physical quantities: predictions for the proton--neutron mass difference, and the hadronic background for measurements of the proton charge radius. All the components of this Compton amplitude can be determined experimentally, except for its subtraction function, $S_1(Q^2)$. Therefore, a first-principles lattice QCD calculation of this subtraction function is of great interest.

\begin{figure}

     \begin{subfigure}{0.5\textwidth}
        \centering
        \scalebox{0.8}{\tikzset{every picture/.style={line width=0.75pt}} 

\begin{tikzpicture}[x=0.75pt,y=0.75pt,yscale=-1,xscale=1]

\draw  [fill={rgb, 255:red, 155; green, 155; blue, 155 }  ,fill opacity=0.59 ] (246,80) .. controls (246,72.82) and (280.03,67) .. (322,67) .. controls (363.97,67) and (398,72.82) .. (398,80) .. controls (398,87.18) and (363.97,93) .. (322,93) .. controls (280.03,93) and (246,87.18) .. (246,80) -- cycle ;
\draw    (171,81) -- (246,80) ;
\draw [shift={(208.5,80.5)}, rotate = 539.24] [fill={rgb, 255:red, 0; green, 0; blue, 0 }  ][line width=0.08]  [draw opacity=0] (8.93,-4.29) -- (0,0) -- (8.93,4.29) -- cycle    ;
\draw    (398,80) -- (473,79) ;
\draw [shift={(435.5,79.5)}, rotate = 539.24] [fill={rgb, 255:red, 0; green, 0; blue, 0 }  ][line width=0.08]  [draw opacity=0] (8.93,-4.29) -- (0,0) -- (8.93,4.29) -- cycle    ;
\draw  [line width=0.75]  (246.2,80.57) .. controls (245.63,78.91) and (244.33,76.73) .. (242.98,74.48) .. controls (240.78,70.83) and (238.69,67.35) .. (239.74,65.99) .. controls (240.79,64.63) and (244.69,65.76) .. (248.78,66.96) .. controls (252.87,68.16) and (256.77,69.3) .. (257.82,67.94) .. controls (258.87,66.58) and (256.78,63.09) .. (254.58,59.44) .. controls (252.39,55.79) and (250.3,52.31) .. (251.35,50.95) .. controls (252.4,49.58) and (256.3,50.72) .. (260.39,51.92) .. controls (264.48,53.12) and (268.38,54.26) .. (269.43,52.9) .. controls (270.48,51.54) and (268.39,48.05) .. (266.19,44.4) .. controls (264,40.75) and (261.91,37.26) .. (262.96,35.9) .. controls (264.01,34.54) and (267.91,35.68) .. (272,36.88) .. controls (275.38,37.87) and (278.62,38.82) .. (280.24,38.35) ;
\draw  [line width=0.75]  (279.72,37.94) .. controls (280.64,36.23) and (281.52,33.44) .. (282.43,30.57) .. controls (283.9,25.9) and (285.31,21.45) .. (286.94,21.47) .. controls (288.57,21.5) and (289.97,25.99) .. (291.43,30.7) .. controls (292.89,35.42) and (294.28,39.91) .. (295.91,39.93) .. controls (297.54,39.96) and (298.95,35.51) .. (300.43,30.84) .. controls (301.9,26.16) and (303.31,21.72) .. (304.94,21.74) .. controls (306.57,21.76) and (307.97,26.25) .. (309.42,30.97) .. controls (310.88,35.69) and (312.28,40.18) .. (313.91,40.2) .. controls (315.54,40.22) and (316.95,35.78) .. (318.42,31.1) .. controls (319.9,26.43) and (321.31,21.98) .. (322.94,22.01) .. controls (324.57,22.03) and (325.96,26.52) .. (327.42,31.24) .. controls (328.88,35.95) and (330.28,40.44) .. (331.91,40.47) .. controls (333.53,40.49) and (334.95,36.04) .. (336.42,31.37) .. controls (337.9,26.7) and (339.31,22.25) .. (340.94,22.27) .. controls (342.57,22.3) and (343.96,26.79) .. (345.42,31.5) .. controls (346.88,36.22) and (348.28,40.71) .. (349.9,40.73) .. controls (351.53,40.76) and (352.94,36.31) .. (354.42,31.64) .. controls (355.9,26.97) and (357.31,22.52) .. (358.94,22.54) .. controls (360.56,22.57) and (361.96,27.05) .. (363.42,31.77) .. controls (364.67,35.82) and (365.88,39.69) .. (367.22,40.73) ;
\draw  [line width=0.75]  (367.83,39.97) .. controls (369.76,39.8) and (372.54,39.02) .. (375.4,38.21) .. controls (380.06,36.89) and (384.5,35.64) .. (385.4,37) .. controls (386.29,38.36) and (383.4,41.95) .. (380.36,45.72) .. controls (377.32,49.49) and (374.42,53.08) .. (375.32,54.44) .. controls (376.22,55.8) and (380.66,54.55) .. (385.32,53.23) .. controls (389.98,51.92) and (394.42,50.67) .. (395.31,52.03) .. controls (396.21,53.38) and (393.32,56.98) .. (390.27,60.74) .. controls (387.23,64.51) and (384.34,68.1) .. (385.24,69.46) .. controls (386.13,70.82) and (390.57,69.57) .. (395.23,68.26) .. controls (399.89,66.94) and (404.33,65.69) .. (405.23,67.05) .. controls (406.13,68.41) and (403.23,72) .. (400.19,75.77) .. controls (399.19,77) and (398.21,78.22) .. (397.37,79.35) ;

\draw (210.5,83.9) node [anchor=north west][inner sep=0.75pt]    {$p$};
\draw (437.5,82.9) node [anchor=north west][inner sep=0.75pt]    {$p$};

\end{tikzpicture}}
        \vspace{3mm}
        \caption{Electromagnetic self-energy}
        \label{fig:2.a}
    \end{subfigure}
    \begin{subfigure}{0.5\textwidth}
        \centering
        \scalebox{0.8}{\tikzset{every picture/.style={line width=0.75pt}} 

\begin{tikzpicture}[x=0.75pt,y=0.75pt,yscale=-1,xscale=1]

\draw  [fill={rgb, 255:red, 155; green, 155; blue, 155 }  ,fill opacity=0.59 ] (246,79) .. controls (246,72.37) and (280.03,67) .. (322,67) .. controls (363.97,67) and (398,72.37) .. (398,79) .. controls (398,85.63) and (363.97,91) .. (322,91) .. controls (280.03,91) and (246,85.63) .. (246,79) -- cycle ;
\draw   (246.26,77.99) .. controls (242.81,76.62) and (239.64,75.29) .. (239.59,73.64) .. controls (239.54,71.92) and (242.89,70.34) .. (246.41,68.68) .. controls (249.93,67.01) and (253.28,65.43) .. (253.22,63.71) .. controls (253.17,61.99) and (249.72,60.62) .. (246.11,59.18) .. controls (242.49,57.74) and (239.05,56.37) .. (238.99,54.65) .. controls (238.94,52.93) and (242.29,51.35) .. (245.81,49.68) .. controls (249.33,48.02) and (252.68,46.44) .. (252.62,44.72) .. controls (252.57,43) and (249.12,41.63) .. (245.51,40.19) .. controls (241.89,38.75) and (238.45,37.38) .. (238.39,35.66) .. controls (238.34,33.94) and (241.69,32.36) .. (245.21,30.69) .. controls (248.73,29.03) and (252.08,27.44) .. (252.02,25.73) .. controls (252.02,25.48) and (251.94,25.24) .. (251.8,25.01) ;
\draw   (398.22,77.96) .. controls (402.22,76.34) and (406.02,74.79) .. (405.99,73.07) .. controls (405.96,71.35) and (402.1,69.94) .. (398.05,68.46) .. controls (394,66.99) and (390.14,65.58) .. (390.11,63.86) .. controls (390.08,62.14) and (393.88,60.59) .. (397.88,58.96) .. controls (401.87,57.34) and (405.68,55.79) .. (405.64,54.07) .. controls (405.61,52.35) and (401.76,50.94) .. (397.71,49.47) .. controls (393.65,47.99) and (389.8,46.58) .. (389.77,44.86) .. controls (389.74,43.14) and (393.54,41.59) .. (397.53,39.97) .. controls (401.53,38.35) and (405.33,36.79) .. (405.3,35.08) .. controls (405.27,33.36) and (401.41,31.95) .. (397.36,30.47) .. controls (393.31,28.99) and (389.45,27.58) .. (389.42,25.86) .. controls (389.41,25.22) and (389.93,24.6) .. (390.8,24) ;
\draw    (171,80) -- (246,79) ;
\draw [shift={(208.5,79.5)}, rotate = 539.24] [fill={rgb, 255:red, 0; green, 0; blue, 0 }  ][line width=0.08]  [draw opacity=0] (8.93,-4.29) -- (0,0) -- (8.93,4.29) -- cycle    ;
\draw    (398,79) -- (473,78) ;
\draw [shift={(435.5,78.5)}, rotate = 539.24] [fill={rgb, 255:red, 0; green, 0; blue, 0 }  ][line width=0.08]  [draw opacity=0] (8.93,-4.29) -- (0,0) -- (8.93,4.29) -- cycle    ;
\draw    (177,26) -- (252,25) ;
\draw [shift={(214.5,25.5)}, rotate = 539.24] [fill={rgb, 255:red, 0; green, 0; blue, 0 }  ][line width=0.08]  [draw opacity=0] (8.93,-4.29) -- (0,0) -- (8.93,4.29) -- cycle    ;
\draw    (392,23) -- (467,22) ;
\draw [shift={(429.5,22.5)}, rotate = 539.24] [fill={rgb, 255:red, 0; green, 0; blue, 0 }  ][line width=0.08]  [draw opacity=0] (8.93,-4.29) -- (0,0) -- (8.93,4.29) -- cycle    ;
\draw    (252,25) -- (392,23) ;
\draw [shift={(322,24)}, rotate = 539.1800000000001] [fill={rgb, 255:red, 0; green, 0; blue, 0 }  ][line width=0.08]  [draw opacity=0] (8.93,-4.29) -- (0,0) -- (8.93,4.29) -- cycle    ;

\draw (190,4.4) node [anchor=north west][inner sep=0.75pt]    {$k$};
\draw (432,2.4) node [anchor=north west][inner sep=0.75pt]    {$k$};
\draw (210.5,82.9) node [anchor=north west][inner sep=0.75pt]    {$p$};
\draw (437.5,81.9) node [anchor=north west][inner sep=0.75pt]    {$p$};
\draw (221,42.4) node [anchor=north west][inner sep=0.75pt]    {$q$};
\draw (409,43.4) node [anchor=north west][inner sep=0.75pt]    {$-q$};

\end{tikzpicture}}
        \caption{Two-photon exchange}
        \label{fig:2.b}
    \end{subfigure}
 
 \caption{Two processes which have dependence on hadronic structure given by the Compton amplitude.}
 
\end{figure}
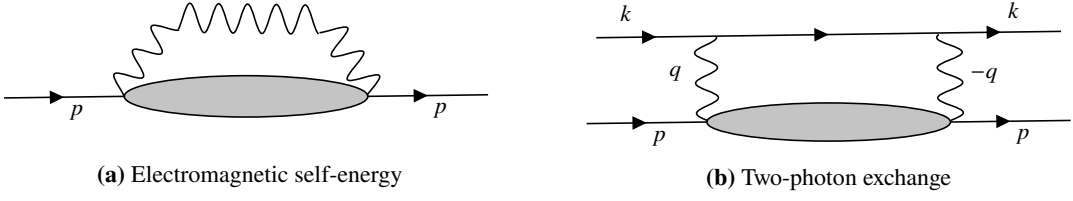

The mass difference of the proton and neutron has two sources: the different masses of the up and down quarks, and the different charges of these quarks---see Refs.~\cite{Gasser:1982ap, Miller:1990iz} for reviews. The leading electromagnetic contribution to the mass difference (Fig.~\ref{fig:2.a}) can be evaluated from the Cottingham sum rule \cite{Cottingham_1963},
\begin{equation}
    \delta m^{\text{EM}} = -\frac{i}{2m_p}\frac{\alpha}{(2\pi)^3} \int d^4 q \frac{T^\mu_\mu(p,q)}{Q^2 -i\epsilon},
    \label{massdiff}
\end{equation}
where $T_{\mu\nu}$ is the forward, spin-averaged Compton amplitude for a proton:
\begin{equation}\label{compdef}
    T_{\mu \nu}(p,q) \equiv \frac{i}{2}\sum_s \int d^4z \  e^{i q\cdot z} \langle p, s | {T} \{ j_\mu(z)j_\nu(0) \} | p, s \rangle.
\end{equation}
This amplitude describes the process of photon-proton scattering, $\gamma^{(*)}(q)P(p) \to \gamma^{(*)}(q)P(p)$, with no momentum transfer between initial and final states (Fig.~\ref{fig:1}).

Similarly, the Compton amplitude is required to constrain measurements of the proton charge radius from the muonic-hydrogen Lamb shift. Crucially, recent determinations of the charge radius from this Lamb shift conflict with previous results obtained via electron--proton scattering \cite{Bernauer_2010} by seven standard deviations \cite{Pohl:2010}---the so-called `proton radius puzzle' \cite{Pohl:2013yb, Antognini:2022xoo}. The hadronic corrections to the Lamb shift are dependent on the two-photon-exchange diagram (Fig.~\ref{fig:2.b}) \cite{Tomalak:2015, Afanasev:2017}: 
\begin{equation}
  \mathcal{M}_{\text{TPE}} = -ie^4 \int \frac{d^4q}{(2\pi)^4}\frac{T_{\mu\nu}L^{\mu\nu}}{Q^4-i\epsilon},
    \label{tpe}
\end{equation}
where $L_{\mu\nu}$ is the leptonic contribution, which can be
calculated from QED, and $T_{\mu\nu}$ is the proton Compton amplitude.
Since the Compton subtraction function is poorly constrained, it
contributes the dominant uncertainty to the hadronic background
\cite{Carlson_2011, Paz_2011, Miller:2012}.
Hence more precise determinations of the subtraction function could
help clarify the proton radius puzzle.

It has been conjectured that the subtraction function receives
contributions from a $J=0$ fixed pole, which is independent of $Q^2$
\cite{Brodsky:1973hm}.
Such a singularity may arise from the exchange of a particle of
spin-zero or from a contact interaction term.
If correct, this would have far-reaching phenomenological consequences \cite{Brodsky:2008qu}, and potentially thwart the Cottingham sum rule.
However, this conjecture has remained rather controversial \cite{Muller:2015vha}.

The starting point for our study is the spin-averaged Compton
amplitude, Eq.~\eqref{compdef}, which can be decomposed into two
structures:
\begin{equation}\label{tensdecomp}
    T_{\mu \nu }(p,q) = \left(-g_{\mu\nu} + \frac{q_\mu q_\nu}{q^2} \right)\mathcal{F}_1(\omega,Q^2) + \left(p_\mu - \frac{p \cdot q}{q^2} q_\mu \right)\left(p_\nu - \frac{p \cdot q}{q^2} q_\nu \right)\frac{\mathcal{F}_2(\omega,Q^2)}{p \cdot q},
\end{equation}
where $\mathcal{F}_{1,2}$ are the \emph{Compton structure functions}, defined in terms of the photon virtuality, $Q^2 = {-q^2}$, and the inverse Bjorken scaling variable, $\omega = 2(p \cdot q)/Q^2$. These structure functions satisfy the following fixed-$Q^2$ dispersion integrals \cite{Drechsel:2002}:
\begin{align}\label{f1}
    \overline{\mathcal{F}}_1(\omega,Q^2) = \mathcal{F}_1(\omega,Q^2) - \mathcal{F}_1(0,Q^2) &= 2\omega^2 \int_0^1 dx \frac{2x F_1(x,Q^2)}{1- x^2\omega^2 - i\epsilon}, \\
    \mathcal{F}_2(\omega,Q^2) &= 4\omega \int_0^1 dx \frac{ F_2(x,Q^2)}{1- x^2\omega^2 - i\epsilon},
    \label{f2}
\end{align}
where $F_{1,2}(x=\omega^{-1}, Q^2)$ are measurable from deep inelastic scattering (DIS) cross sections.

However, equation \eqref{f1} features a contribution from the \emph{Compton amplitude subtraction function},
\begin{equation}
    S_1(Q^2) \equiv \mathcal{F}_1(\omega =0,Q^2),
\end{equation}
which is not experimentally accessible. Instead, this subtraction function has been determined from model-dependent, dispersive, non-relativistic, and effective theory calculations \cite{Alarcon:2013cba, Caprini:2016wvy, Caprini:2021btw, Carlson_2011, Erben:2014, Gasser:1974, Miller:2012, Paz_2011, Thomas:2014, Tomalak:2015aoa, Tomalak:2018, Walker-Loud_2012, Walker-Loud:2019}. These calculations have sizeable errors and are not always consistent with one another \cite{Gasser_2020}. At large $Q^2$, the subtraction function can be evaluated model-independently using the operator product expansion (OPE) \cite{Collins_1979, Hill_2017}, and the following asymptotic behaviour is predicted:
\begin{equation}
    S_1(Q^2) \sim \frac{m_N^2}{Q^2}, \quad \text{for}\quad Q^2\gg m_N^2.
    \label{opepred}
\end{equation}
However, as can be seen in Eqs.~\eqref{massdiff} and \eqref{tpe}, a determination for the whole domain of $S_1(Q^2)$ is necessary for inputs into the aforementioned physical quantities. Therefore, a determination of this subtraction function, particularly for low and intermediate $Q^2$, is of great interest.

Recently, the Feynman-Hellmann method has been used to calculate the forward Compton amplitude in lattice QCD \cite{Chambers_2017, Utku_2020}, and has been extended to off-forward kinematics \cite{hannafordgunn2021generalised}. In this report, we apply this technique to calculate the Compton subtraction function. In particular, our initial results display large deviations from the asymptotic behaviour predicted by the OPE, Eq.~\eqref{opepred}. Therefore, we conduct an investigation of lattice artifacts of the subtraction function. We vary the lattice volume and spacing, and compare different discretisations of the vector current. Finally, we present a novel method to explore the effects of current-current contact terms in Feynman-Hellmann methods. Hence this work lays the foundation for future lattice QCD calculations of the Compton subtraction function, which would allow for better theoretical predictions of the proton--neutron mass difference, and more precise experimental determinations of the proton charge radius.

\section{Feynman-Hellmann: Local vector current implementation}
\label{sec:local}

Feynman-Hellmann methods provide a feasible alternative to the direct calculation of four-point functions. A background field is introduced to the quark propagator, with a small unphysical coupling, $\lambda$. At order $\lambda^2$, the perturbation to the propagator is a four-point function with a sum over time-slices on which the background field is inserted. As such, one set of Feynman-Hellmann inversions yields the sum over time-slices necessary to calculate a discretisation of the Compton amplitude, Eq.~\eqref{compdef}. By contrast, a direct four-point function evaluation of this amplitude would require $\mathcal{O}(T^2)$ inversions to get all possible insertion times for a lattice of temporal extent $T$.

It was shown in Ref.~\cite{Utku_2020} that for a perturbed nucleon propagator, $\mathcal{G}_{\lambda}(t) \simeq A_{\lambda}e^{-E_{\lambda}t}$, the perturbed energy, $E_{\lambda}$, can be related to the Compton amplitude subtraction function:
\begin{equation}
  - m_N \frac{\partial^2 E_{\lambda}}{\partial \lambda^2 }\bigg |_{\lambda=0} = S_1(Q^2),
    \label{fh_relation}
\end{equation}
in the nucleon's rest frame: $\vec{p}=0$. The derivation of this Feynman-Hellmann relation, Eq.~\eqref{fh_relation}, was carried out in terms of hadronic states with continuous spacetime coordinates \cite{Utku_2020}. Since our aim here will be on understanding and controlling several lattice artifacts, we instead focus on the Feynman-Hellmann implementation at the level of lattice quark propagators.

A common discretisation of the electromagnetic current is the \emph{local current}:
\begin{equation}
    j_{\mu}^{\text{loc}} (n) = \bar{\psi}(n)\gamma_{\mu}\psi(n).
    \label{loccurrent}
\end{equation}
To calculate the Compton amplitude, Eq.~\eqref{compdef}, with this discretisation, we compute the following perturbed propagator:
\begin{equation}
    \begin{split}
        S_{{\lambda}} = \big[M- \lambda \mathcal{O}_{\vec{q}}\big]^{-1},
      \label{perturbedquarkprop}
    \end{split}
\end{equation}
where the perturbing matrix is
\begin{equation}
    \big[\mathcal{O}_{\vec{q}}\big]_{n,m} =  \delta_{n,m} \phi_{\vec{q}}(\vec{n})i\gamma_3, \quad \text{with} \quad \phi_{\vec{q}}(\vec{n}) = e^{i\vec{q}\cdot\vec{n}}+e^{-i\vec{q}\cdot\vec{n}},
    \label{localcurrent}
\end{equation}
for $\lambda$ a small coupling. $M$ is the fermion matrix, which can be written, up to the clover term \cite{clover}, as
\begin{equation}
    M(n,m) \equiv \left(m_0+4\right )\delta_{n,m} - \frac{1}{2a}\sum_{\mu}\bigg [\vec{\mathcal{D}}_{\mu}(n,m) + \cev{\mathcal{D}}_{\mu}(n,m) \bigg],
\end{equation}
with
\begin{equation*}
    \vec{\mathcal{D}}_{\mu}(n,m)\equiv (1-\gamma_{\mu})U_{\mu}(n)\delta_{n+\hat{\mu},m}, \quad \cev{\mathcal{D}}_{\mu}(n,m)\equiv (1+\gamma_{\mu})U_{\mu}(n-\hat{\mu})^{\dagger} \delta_{n-\hat{\mu},m}.
\end{equation*}
Therefore, the second derivative of the perturbed propagator is
\begin{equation}
     \frac{\partial^2}{\partial \lambda^2}{S_{{\lambda}}}\bigg|_{{\lambda}=0}  = 2M^{-1}\mathcal{O}_{{\vec{q}}} M^{-1}\mathcal{O}_{{\vec{q}}} M^{-1},
     \label{localsecondderiv}
\end{equation}
which is a four-point function with the {local current}, Eq.~\eqref{loccurrent}.

In terms of the quark propagators, the nucleon propagators with the perturbation to the $u$ or $d$ quark\footnote{Since the perturbation is only applied to the fermion propagators and not to the sea quarks, these and all other results in this report are connected only.} are written, up to spin/colour structure, as
\begin{equation}
    \mathcal{G}^{uu}_{\lambda} \sim \big\< S^u_{\lambda}S^u_{\lambda}S^d \big\>, \quad  \mathcal{G}^{dd}_{\lambda} \sim \big\< S^uS^uS^d_{\lambda} \big\>, \quad  \mathcal{G}^{ud}_{\lambda_1,\lambda_2} \sim \big\< S^u_{\lambda_1}S^u_{\lambda_1}S^d_{\lambda_2} \big\>.
\end{equation}
Recalling the spectral representation of the nucleon perturbed propagator, $\mathcal{G}^{qq'}_{\lambda}(t) \simeq A^{qq'}_{\lambda}e^{-E^{qq'}_{\lambda}t}$, we have the following flavour-dependent Feynman-Hellmann relations:
\begin{equation}
     -m_N \frac{\partial^2 E^{qq}_{\lambda}}{\partial \lambda^2 }\bigg |_{\lambda=0} = S^{qq}_1(Q^2), \quad -2m_N \frac{\partial^2 E^{ud}_{\lambda}}{\partial \lambda_1\partial\lambda_2 }\bigg |_{\lambda_1=\lambda_2=0} = S^{ud}_1(Q^2)+S_1^{du}(Q^2).
\end{equation}
Therefore, to calculate the proton contribution to the subtraction function, we take
\begin{equation}
   S_1^p(Q^2) = \sum_{q,q' = u,d}e_{q}e_{q'}S^{qq'}_1(Q^2) = \frac{4}{9} S_1^{uu}(Q^2) + \frac{1}{9}S_1^{dd}(Q^2)-\frac{2}{9}\big(S_1^{ud}(Q^2)+S_1^{du}(Q^2)\big).
\end{equation}

\begin{table}
    \caption{Gauge ensemble details}
    \centering
\[
\begin{array}{c c c c c c c c c}
    N_f & L^3 \times T & L  [\text{fm}]  & a [\text{fm}]& \beta &\kappa  & m_{\pi} [\text{GeV}] & Z_V   \\ 
    \hline \hline
    2+1 & 32^3 \times 64 & 2.4 &  0.074 & 5.50  & 0.120900  & 0.47 & 0.86  \\  
      \hline
     2+1 & 48^3 \times 96  & 3.3 & 0.068 & 5.65 & 0.122005 & 0.41 & 0.87 \\ 
     & 48^3 \times 96 &2.8  &0.058 & 5.80 & 0.122810 & 0.43  & 0.88 \\  

\end{array}
\]
    \label{tab:I}
\end{table}

\subsection*{Results: Local Current Implementation}

The simulations with the local current implementation were carried out on three different gauge ensembles generated by the QCDSF/UKQCD Collaborations \cite{configs} (Tab.~\ref{tab:I}), with varying volume, lattice spacing and quark masses. All three ensembles are at the SU(3) flavour symmetric point: $\kappa_l = \kappa_s=\kappa$. The inserted momentum is always chosen to be of the form $q_{\mu} = (0, \vec{q})$. Therefore, to calculate the subtraction function, for which $p\cdot q=0$, we simply choose our nucleon sink momentum $\vec{p} = 0$. For the $32^3 \times 64$ lattice, we calculate a large range of $Q^2$ values, while for the two larger volumes, there are fewer. For the majority of points, the statistics are $N_{\text{meas}} \sim \mathcal{O}(100)- \mathcal{O}(1000)$, with some higher statistics points, $N_{\text{meas}} \sim \mathcal{O}(10000)$, for the smallest volume.



\begin{figure}
    \centering
    \includegraphics[width =0.76\textwidth, keepaspectratio]{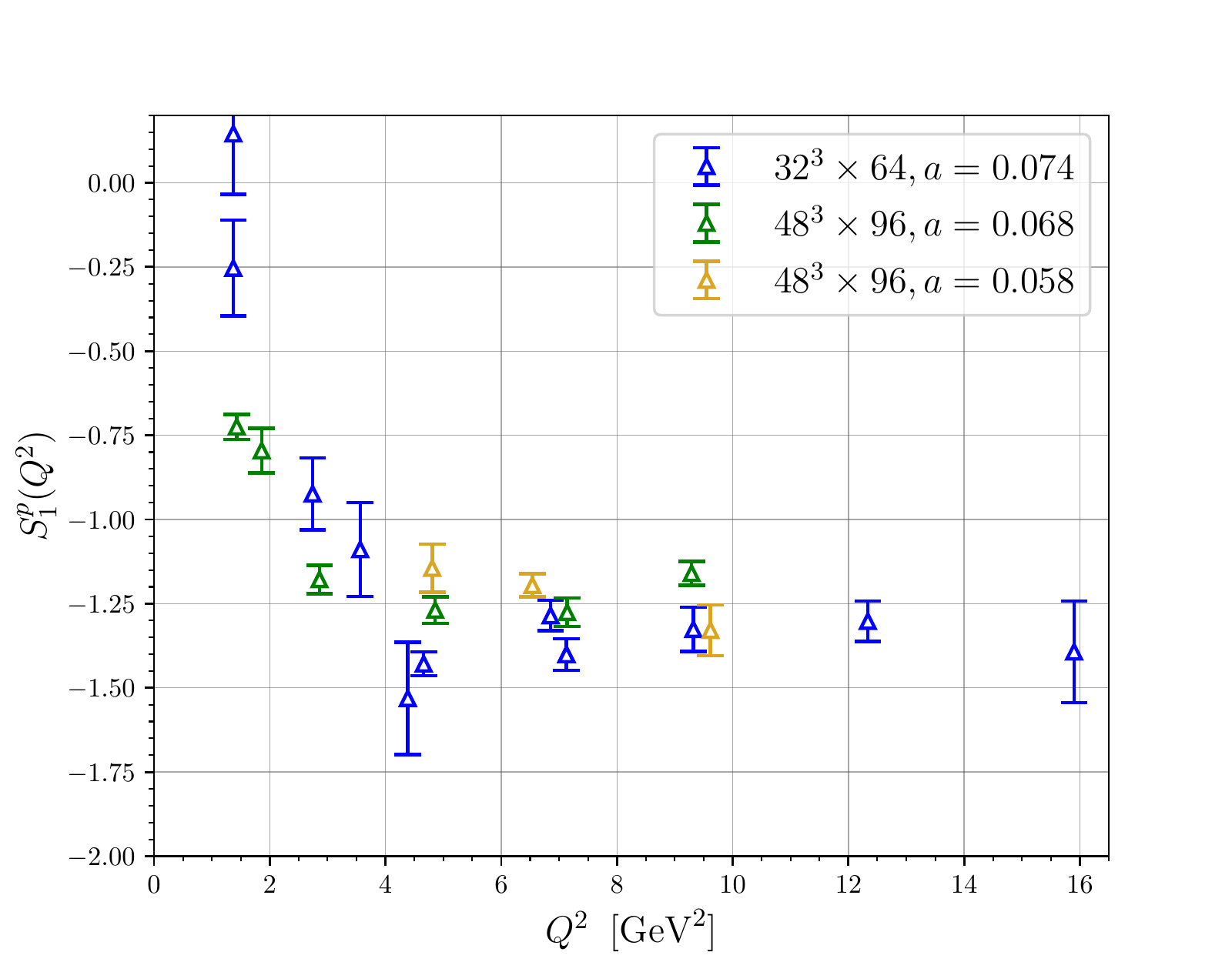}
    \caption{Local current proton subtraction function results for a range of lattices. }
     \label{fig:local_results}
\end{figure}

The results are presented in Fig.~\ref{fig:local_results}, and show an asymptotic behaviour that clearly deviates from the OPE prediction of $S_1(Q^2)\sim Q^{-2}$, given in Eq.~\eqref{opepred}. Instead of trending to zero for $Q^2\gg m_N^2$, the subtraction function approaches a non-zero value.

In Regge analysis, it has been pointed out that the Compton amplitude may contain an OPE-breaking `fixed pole' \cite{Creutz:1968, Brodsky:2008qu}, which could possibly account for the results in Fig.~\ref{fig:local_results}. 

However, since the OPE is such a successful tool, we must investigate whether or not this anomalous asymptotic behaviour is due to a lattice artifact.
As such artifacts will vanish in the continuum limit, we might expect them to be sensitive to variations in the lattice volume and spacing. However, the results in Fig.~\ref{fig:local_results} indicate only very minimal volume and spacing dependence in the subtraction function. This is in agreement with recent calculations using baryon chiral perturbation theory, where it was found that finite volume corrections to the Compton amplitude subtraction function would indeed be small \cite{Lozano_2021}.

Since the anomalous asymptotic behaviour of the subtraction function does not vary greatly with changes in volume and spacing, we next investigate how it depends on the discretisation of the vector current.

\section{Feynman-Hellmann: Conserved vector current implementation}
\label{sec:conserved}

The results shown in the previous section were based on the \emph{local} discretisation of the vector current, Eq.~\eqref{loccurrent}. In this section, we will repeat the same calculation with the \emph{conserved} vector current,
\begin{equation}
   \begin{split}
        j^{\text{con}}_{\mu}(n) = \frac{1}{2}\bar{\psi}(n)\left(\vec{\mathcal{D}}_\mu(n,m) - \cev{\mathcal{D}}_\mu(n,m) \right) \psi(m).
   \end{split}
\end{equation}
For the Wilson fermion action, this operator is a Noether current, with a renormalisation factor of $Z_V=1$, in contrast to the local operator. However, our implementation of this current followed here introduces an unphysical contamination to the energy shift, which we refer to as the \emph{seagull term}. 

We implement the conserved current by introducing a perturbation on the gauge links: 
\begin{equation}
    U_{\mu}(n) \to [1 + \delta_{\mu 3} ( e^{i\lambda \phi(n)} - 1 )] U_{\mu}(n).
    \label{pertlinks}
\end{equation}
Note that different implementations are possible; however, by modifying the gauge links, we can use existing optimised algorithms to more efficiently invert the fermion matrix.

\begin{figure}
    \centering
    \includegraphics[width =0.76\textwidth, keepaspectratio]{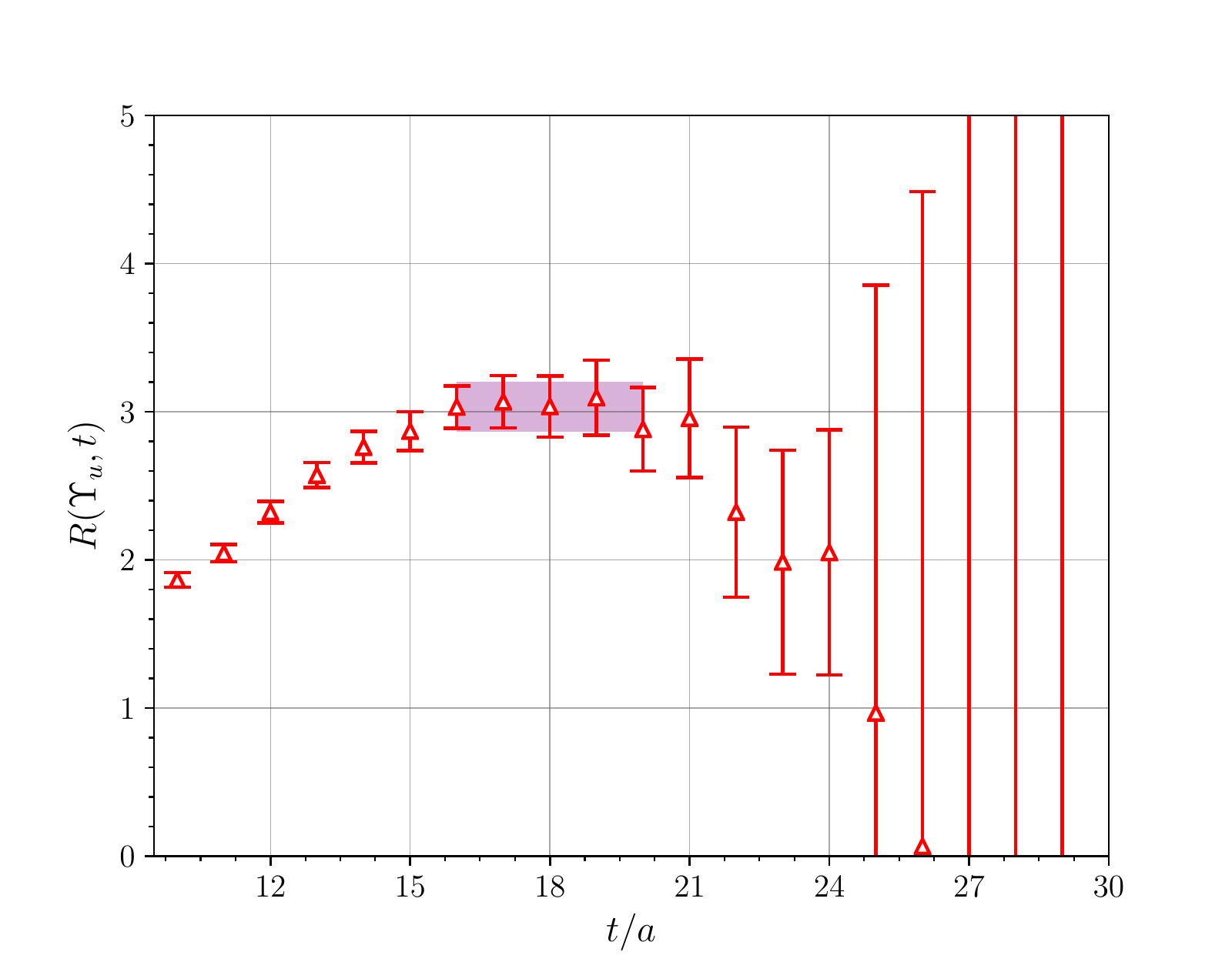}
    \caption{Plot of the ratio $R(\Upsilon_u, t) = \mathcal{G}_3(\Upsilon_u,t)/\mathcal{G}_2(t)$ where the baryon three-point function was calculated with fixed insertion at $t=10$. The purple shaded horizontal region is the fitting window ($t \in [16,20]$). }
    \label{fig:4}
\end{figure}

Therefore, the perturbed fermion matrix is
\begin{equation}
    M_{\lambda}(n,m) =  \left(m_0 + 4\right)\delta_{n,m} -\frac{1}{2a} \sum_{\mu}\bigg[ \vec{\mathcal{D}}_{\mu}(n,m,\lambda)+ \cev{\mathcal{D}}_{\mu}(n,m,\lambda) \bigg],
\end{equation}
with
\begin{align*}
    \vec{\mathcal{D}}_{\mu}(n,m,\lambda) &\to  \vec{\mathcal{D}}_{\mu}(n,m) \left[1 + \delta_{\mu 3} \left( e^{i\lambda \phi(n)} - 1 \right)\right], \\
    \cev{\mathcal{D}}_{\mu}(n,m,\lambda) &\to  \cev{\mathcal{D}}_{\mu}(n,m) \left[1 + \delta_{\mu 3} \left( e^{-i\lambda \phi(n+\hat{\mu})} - 1 \right)\right].
\end{align*}
 \begin{figure}[h]
    \centering
     \includegraphics[width =0.76\textwidth, keepaspectratio]{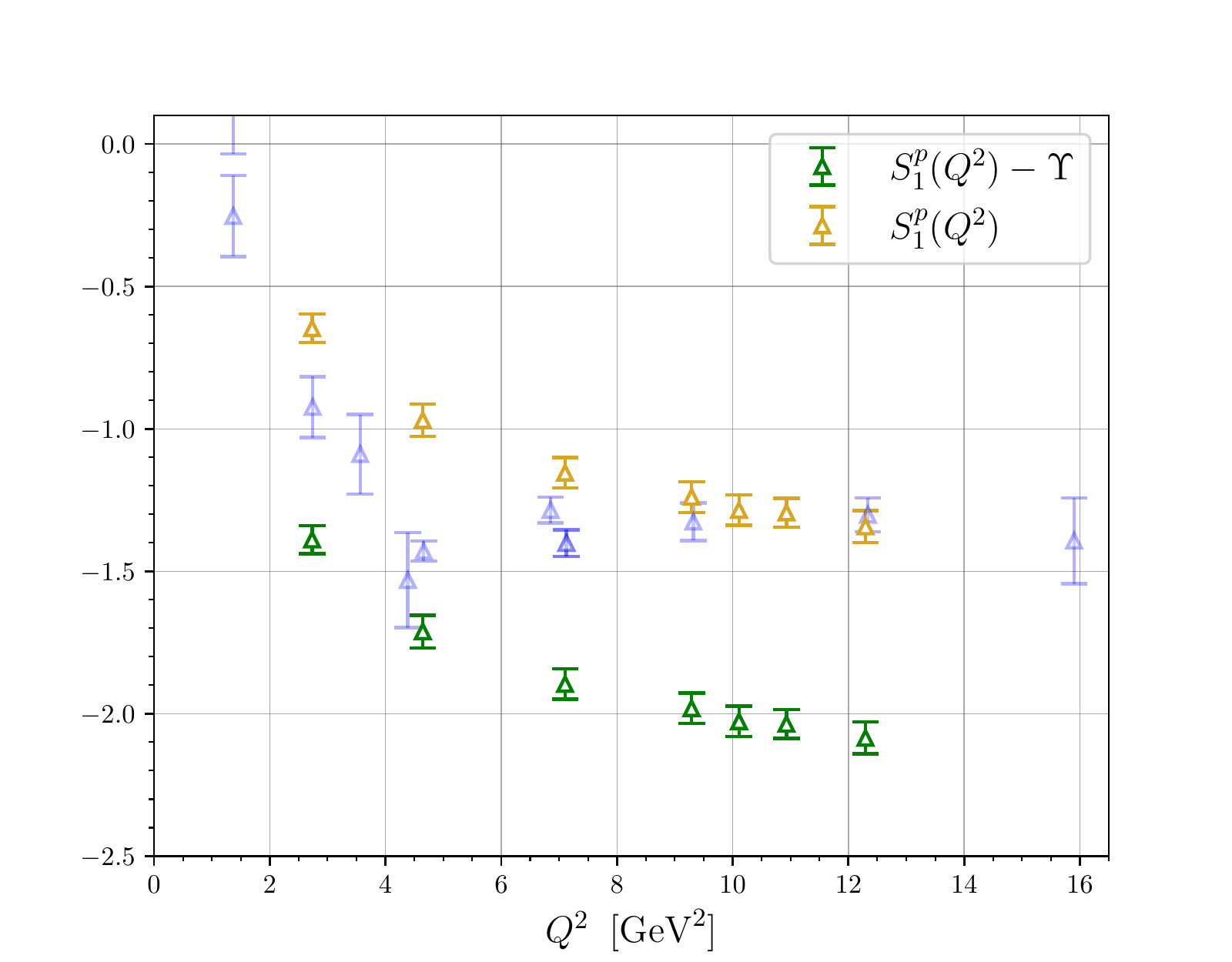}
     \caption{Proton subtraction function results using the conserved current implementation, before (green) and after (yellow) removal of seagull term, $\Upsilon$. The results in blue are from the local implementation seen in Fig. \ref{fig:local_results}.}
     \label{fig:conserved_results}
 \end{figure}
The perturbed propagator is then $S_{\lambda} = [M_{\lambda}]^{-1}$, which has the second derivative
\begin{equation}
     \frac{\partial^2}{\partial \lambda^2}{S_{{\lambda}}}\bigg|_{{\lambda}=0}  = \underbrace{M^{-1}\big(\phi_{\vec{q}}\vec{\mathcal{D}}_3+\cev{\mathcal{D}}_3 \phi_{\vec{q}}\big) M^{-1}}_{\text{seagull contribution}} + \underbrace{2M^{-1}\big(\phi_{\vec{q}}\vec{\mathcal{D}}_3-\cev{\mathcal{D}}_3 \phi_{\vec{q}} \big) M^{-1}\big(\phi_{\vec{q}}\vec{\mathcal{D}}_3-\cev{\mathcal{D}}_3\phi_{\vec{q}} \big) M^{-1}}_{\text{four-point function}}.
     \label{conservedsecondderiv}
\end{equation}
Therefore, the Feynman-Hellmann relation, Eq.~\eqref{fh_relation}, has an additional \emph{seagull term} for this conserved current implementation:
\begin{equation}
  - m_N \frac{\partial^2 E_{\lambda}}{\partial \lambda^2 }\bigg |_{\lambda=0} = S_1(Q^2) - \Upsilon,
    \label{fh_relation_seagull}
\end{equation}
where the seagull term is given by
\begin{equation}
   \Upsilon_q = 2\<N|\bar{\psi}_q(n)\big(\vec{\mathcal{D}}_3+\cev{\mathcal{D}}_3 \big)\psi_q(n)|N\>.
    \label{seagull}
\end{equation}
Note that $(\phi_{\vec{q}})^2 = 2 + e^{2i\vec{q}\cdot\vec{n}}+e^{-2i\vec{q}\cdot\vec{n}} $, and the $e^{\pm 2i\vec{q}\cdot\vec{n}}$ terms do not produce a contribution to the energy shift. Hence $\Upsilon$ is independent of the transferred momentum $Q^2$.

\subsection{Seagull results}
As discussed in the preceding section, the seagull term arises from the fact that the perturbation to the fermion matrix is not a simple linear shift, but was instead applied to the gauge links, Eq.~\eqref{pertlinks}. Since this seagull term is purely an artifact of the implementation process, we must remove it from our calculation of the energy shift to isolate the subtraction function.

The three-point function corresponding to the seagull term is computed using a conventional sequential source through the operator. The required matrix element is then extracted from the ratio of the three-point and two-point functions:
\begin{equation}
   R(\Upsilon_q, t)  = \frac{\mathcal{G}_3(\Upsilon_q, t)}{\mathcal{G}_2(t)} 
   \overset{t\gg a}{\simeq}\Upsilon_q,
\end{equation}
for $\Upsilon_q$ as defined in Eq.~\eqref{seagull}. The results for the up-quark are presented in Fig.~\ref{fig:4}. Once this matrix element is calculated we remove it from the conserved current calculation, as per Eq.~\eqref{fh_relation_seagull}, to isolate the subtraction function.

\subsection{Results: Conserved current implementation}

The proton subtraction function is calculated for the conserved current using the same gauge ensemble as the $32^3 \times 64$ volume local current seen in Tab. \ref{tab:I} with $N_{\text{meas}} = 1000$.

In comparison to the local current, we can see in Fig.~\ref{fig:conserved_results} that the conserved current subtraction function exhibits different behaviour in its intermediate $Q^2$ values. Once we remove the  seagull term to isolate the subtraction function, the large $Q^2$ behaviour closely matches OPE-breaking behaviour of the local current. Therefore, the anomalous asymptotic behaviour of the subtraction function can not be attributed directly to the discretisation of the current.

\section{Temporal Interlacing}
\label{sec:interlacing}

As outlined in the previous sections, the observed anomalous
asymptotic behaviour of $S_1(Q^2)$ changes very little with different
discretisations of the current or with the variations in lattice
spacing and volume.
Here, we explore a Feynman-Hellmann implementation that allows
us to sample the integration region of the Compton amplitude,
Eq.~\eqref{compdef}, more coarsely.
The motivation for considering such an implementation is that we will
remove any contamination from a potential contact term, however it
comes with the caveat that this method will also remove any essential
short-distance contributions, such as the $Z$-graph which is proposed
to be responsible for the elusive fixed pole \cite{Brodsky:1973hm}.
Keeping this in mind, however, by exploring such an implementation we
hope to gain insights into whether or not the observed anomalous
behaviour is due to a short-distance effect.

As previously discussed, our calculation has a sum over all time
slices on which the currents are inserted, including contributions for
which the Euclidean separation of currents is $|z|\sim a$, with
$z_{\mu}$ as in Eq.~\eqref{compdef}.
As a first step for investigating the effects of these contributions,
we implement a coarser sampling of the temporal integration region by
inserting two currents on different sets of time slices thereby
introducing a minimum temporal separation, $\tmin$.
%
For instance, in the simplest case where $\tmin = 1$, we insert one
current on the even time slices and the other on the odd:
%
\begin{equation}
   [\mathcal{O}_1]_{n,m} = \delta^{\text{even}}_{t_n,t_m}\delta_{\vec{n}, \vec{m}}\phi_{\vec{q}}(\vec{n})i\gamma_3,\quad   [\mathcal{O}_2]_{n,m} = \delta^{\text{odd}}_{t_n,t_m}\delta_{\vec{n}, \vec{m}}\phi_{\vec{q}}(\vec{n})i\gamma_3,
\end{equation}
where we have defined $\delta^{\text{even}(\text{odd})}$ to be non-vanishing only on even (odd) timeslices.
%
%
The perturbed propagator is now
\begin{equation}
    \begin{split}
        S_{\vec{\lambda}}(z_n;z_m)  & = \bigg[\big[M- \lambda_1 \mathcal{O}_1-\lambda_2 \mathcal{O}_2\big]^{-1}\bigg]_{n,m}.
        \label{perturbedffcorrIL}
    \end{split}
\end{equation}
%
Therefore, the Feynman-Hellmann relation for $\tmin=1$ interlacing is
\begin{equation}
        \begin{split}
         - E_N \frac{\partial^2 E_{\vec{\lambda}}}{\partial \lambda_1\lambda_2 }\bigg |_{\vec{\lambda}=0} =   \sum_{t_1=0,2,4,6...}\sum_{t_2=1,3,5,...}\sum_{\vec{z}}e^{-i\vec{q}\cdot\vec{z}}\<N(\vec{p})|T\{j_{3}(\vec{z},t_1)j_{3}(0, t_2)\}|N(\vec{p})\>.
            \label{interlacingfhrel}
        \end{split}
    \end{equation}
With a judicious choice of kinematics, the RHS of
Eq.~\eqref{interlacingfhrel} is proportional to a discretisation of
$S_1(Q^2)$.
The interlacing in Eq.~\eqref{interlacingfhrel} changes the measure of
the two sums over time-slices from $a\to 2a$, which must be accounted
for by a factor of four. However, once this normalisation is accounted
for, in the continuum limit Eq.~\eqref{interlacingfhrel} approaches
the same object as our previous discretisations of the Compton
amplitude.
Similar results can be derived for $\tmin=2$.


  \begin{figure}
      \centering
    \includegraphics[width=\linewidth]{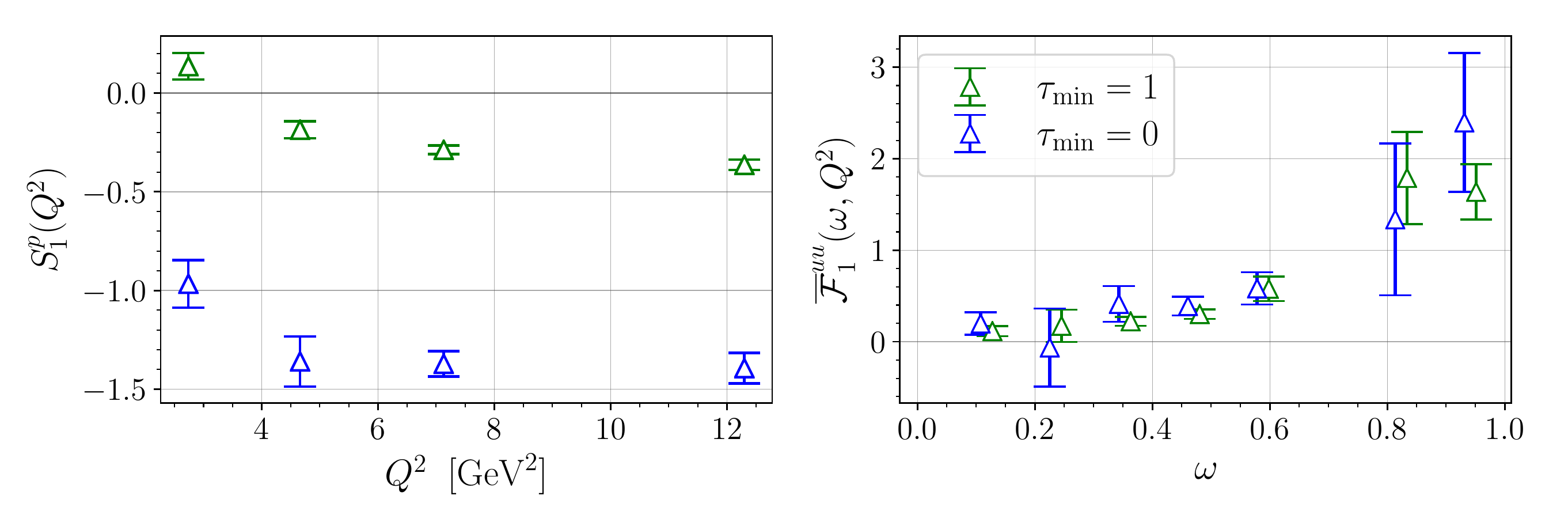}
    \caption{Results for multiple interlacings with the local current.
      Interlacing results have low statistics, $N_{\text{meas}}\approx
      200$, while the uninterlaced ($\tmin=0$) use the implementation
      from section \ref{sec:local}, and have $N_{\text{meas}}\approx
      1000$.
      Left: The proton subtraction function, with OPE prediction from
      Ref.~\cite{Hill_2017}.
      Right: The up quark contribution to the subtracted Compton
      structure function, $\overline{\mathcal{F}}_1$, for $Q^2\approx
      4.7\;\text{GeV}^2$.}
    \label{interlacingres}
  \end{figure}

\subsection{Interlacing results}

Our preliminary results for $S_1(Q^2)$ with the interlacing method are
calculated on the $32^3\times 64$ gauge configurations (see
Tab.~\ref{tab:I}), using the local current, Eq.~\eqref{loccurrent}.
We implement interlacings for two different values of minimum time
separation: $\tmin=0$, which are simply the results from section
\ref{sec:local}, and $\tmin = 1$, for which the new method has been
applied.
Since our interlaced results are exploratory, they have relatively low
statistics: $N_{\text{meas}}\approx200$.

In left plot of Fig.~\ref{interlacingres}, we show the subtraction function with and without interlacing.
We observe that the anomalous asymptotic behaviour of the subtraction
function is reduced in the interlaced results.
We also observe that the $Q^2$ dependence of the two results are
remarkably similar.
Hence it would appear that by removing the $t=0$ contribution to the
integral, we have essentially removed a contribution that is constant
in $Q^2$.
Whether this is due to an unphysical contact term or a fixed pole
remains a question to be addressed in future work.
By contrast, in the right plot of Fig.~\ref{interlacingres}, we
observe that the subtracted Compton structure function,
$\overline{\mathcal{F}}_1$ defined in Eq.~\eqref{f1}, which is
independent of $S_1(Q^2)$, is largely unaffected by the interlacing.

This demonstrates that the anomalous asymptotic behaviour of the
subtraction function can be attributed to very short-distance
contributions (i.e.~$|z|\sim a$), which are removed by the interlacing
procedure.
However, the $|z|\sim a$ contributions apparently do not affect the
Compton structure function $\overline{\mathcal{F}}_1$ significantly.
These very short-distance contributions could be lattice artifacts as
suggested by Martinelli et al. \cite{Dawson:1997, Martinelli_1999}, or
they could be of a physical origin, such as the proposed interactions
giving rise to an OPE-breaking `fixed pole'. However, further
investigation, numerical and analytic, is needed before we draw any
strong conclusions.

\section{Conclusion}

In this report, we present several calculations of the Compton amplitude subtraction function, $S_1(Q^2)$, in lattice QCD.
In contrast to the OPE prediction of $S_1(Q^2) \sim Q^{-2}$, our initial results trend to a large non-zero value at high-energies.
This anomalous behaviour was found to persist even after varying the lattice spacing and changing the current discretisation.

In the final section, we present a novel method, temporal interlacing, that allows us to more coarsely sample the integration region of the Compton amplitude.
Using this method, we demonstrate that the anomalous behaviour of the subtraction function can be attributed to very short-distance contributions, which may be lattice artifacts or physical contributions---future work will aim to clarify this.
We are currently performing an investigation of the Compton amplitude and its subtraction function on configurations with gradient flow as an extension of \cite{Can:2021swa}.
This investigation is ongoing and results will be reported elsewhere.

This work provides a foundation for future lattice QCD calculations of the Compton amplitude subtraction function.
This will allow us to reduce the theoretical uncertainties in predictions for the proton--neutron mass difference and improve determinations of the proton charge radius.

\section{Acknowledgements}

The numerical configuration generation (using the BQCD lattice QCD program~\cite{Haar:2017ubh})) and data analysis (using the Chroma software library~\cite{Edwards:2004sx}) was carried out on the DiRAC Blue Gene Q and Extreme Scaling (EPCC, Edinburgh, UK) and Data Intensive (Cambridge, UK) services, the GCS supercomputers JUQUEEN and JUWELS (NIC, Jülich, Germany) and resources provided by HLRN (The North-German Supercomputer Alliance), the NCI National Facility in Canberra, Australia (supported by the Australian Commonwealth Government) and the Phoenix HPC service (University of Adelaide). AHG is supported by an Australian Government Research Training Program (RTP) Scholarship. RH is supported by STFC through grant ST/P000630/1. PELR is supported in part by the STFC under contract ST/G00062X/1. GS is supported by DFG Grant No. SCHI 179/8-1. KUC, RDY and JMZ are supported by the Australian Research Council grants DP190100297 and DP220103098.

\bibliographystyle{JHEP}
\bibliography{refer}

\end{document}